\documentclass[12pt]{article}
\usepackage{graphics}
\usepackage{epsfig}
\usepackage{subfigure}
\usepackage{graphics}
\usepackage{amsmath}
\usepackage{amsfonts}
\usepackage{amssymb}
\usepackage{url}
\usepackage{hyperref}
\begin{document}
%%%%%%%%%%%%%%%%%%%%%%%%%%%%
\begin{titlepage}
%%%%% PREPRINT NUMBERS %%%%%%

%%%%%%%%%%%%%%%%%%%%%%%%%%%%%%
\vspace{2\baselineskip}
%%%%%%%%%%%%%%%%%%% TITLE %%%%%%%%%%%%%%%%%%
\begin{center}
{\Large\bf 
Flux trapping in superconducting accelerating cavities during cooling down with a spatial temperature gradient
}
\end{center}
%%%%%%%%%%%%%%%% AUTHORS %%%%%%%%%%%%%%%%%%%%%%%
\vspace{0.2cm}
\begin{center}
{\large
Takayuki Kubo%$^{\,\,a}$
%\footnote{\tt E-mail:kubotaka@post.kek.jp}
}
\end{center}
%%%%%%%%%%%%%%%%%%%%%%% AFFILIATION %%%%%%%%%%%%
\vspace{0.2cm}
\begin{center}
%${}^{a}$ 
{KEK, High Energy Accelerator Research Organization, Tsukuba, Ibaraki, Japan.\\
SOKENDAI, The Graduate University for Advanced Studies, Hayama, Kanagawa, Japan}
\end{center}
%\vskip 5mm

\begin{abstract}%
During the cool-down of a superconducting accelerating cavity, 
a magnetic flux is trapped as quantized vortices, 
which yield additional dissipation and contribute to the residual resistance. 
Recently, cooling down with a large spatial temperature gradient attracts much attention for successful reductions of trapped vortices. 
The purpose of the present paper is to propose a model to explain the observed efficient flux expulsions and the role of spatial temperature gradient during the cool-down of cavity. 
In the vicinity of a region with a temperature close to the critical temperature $T_c$,
the critical fields are strongly suppressed and can be smaller than the ambient magnetic field. 
A region with a lower critical field smaller than the ambient field is in the vortex state. 
As a material is cooled down, a region with a temperature close $T_c$ associating the vortex state domain sweeps and passes through the material. 
In this process, vortices contained in the vortex state domain are trapped by pinning centers that randomly distribute in the material. 
A number of trapped vortices can be naively estimated by using the analogy with the beam-target collision event. 
Based on this result, the residual resistance is evaluated. 
We find that a number of trapped vortices and the residual resistance are proportional to the strength of the ambient magnetic field and the inverse of the temperature gradient.  
The obtained residual resistance agrees well with experimental results. 
A material property dependence of a number of trapped vortices is also discussed. 
\end{abstract}

\end{titlepage}

%\subjectindex{superconducting radio-frequency cavity, field limit}

%\maketitle

%\tableofcontents

%%%%%%%%%%%%%%%%%%%%%%
%%%%%%%%%%%%%%%%%%%%%%
\section{Introduction}
%%%%%%%%%%%%%%%%%%%%%%
%%%%%%%%%%%%%%%%%%%%%%

The superconducting radio-frequency (SRF) cavity~\cite{Hasan, gurevich_review} is one of the core components of modern particle accelerators such as the proposed International Linear Collider (ILC)~\cite{TDR}. 
The surface resistance, $R_s$, defines the dissipation at the inner surface of the SRF cavity.
$R_s$ is usually described by a summation of a strongly and a weakly temperature-dependent terms. 
The former and the latter are called the Bardeen-Cooper-Schrieffer (BCS) resistance $R_{\rm BCS}$ and the residual resistance $R_{\rm res}$, respectively. 
The origin of $R_{\rm BCS}$ is the microwave absorptions by thermally excited quasi-particles. 
$R_{\rm BCS}$ for 1.3 GHz Nb cavity at the operating temperature of the ILC, $T\simeq 2\,{\rm K}$, is given by $\mathcal{O}(1) \!- \! \mathcal{O}(10)\,{\rm n\Omega}$, 
depending on material properties. 
As $T\to 0$, quasi-particles ceases to be excited, 
and $R_{\rm BCS}$ vanishes. 
The other term, $R_{\rm res}$, weakly depends on $T$ and remains finite at $T\to 0$.  
It has various origins~\cite{gurevich_ciovati}: 
contributions from a magnetic flux trapped as quantized vortices, 
normal conducting precipitates, a finite density of subgap states, and so on. 
The first one is thought to be the major contribution to $R_{\rm res}$ in the present SRF technology~\cite{gurevich_ciovati, gurevich_TTC2015, vallet, benvenuti, ono, saito}. 
A typical $R_{\rm res}$ observed in present 1.3 GHz TESLA shape Nb cavities is $\mathcal{O}(1) \!-\! \mathcal{O}(10)\,{\rm n\Omega}$, 
which can account for some fraction of $R_s$ for the case of $R_{\rm BCS}\lesssim \mathcal{O}(10)\,{\rm n\Omega}$. 
In particular, impurity doped Nb cavities~\cite{dhakal_IPAC12, dhakal, grassellino} have so small $R_{\rm BCS}$ that $R_{\rm BCS}(2\,{\rm K}) \sim \mathcal{O}(1)\,{\rm n\Omega}$ and $R_{\rm res}$ can dominate $R_s$ even at $T\simeq 2\,{\rm K}$~\cite{umemori_TTC2015}. 
A reduction of a number of trapped vortices has become increasingly important as $R_{\rm BCS}$ decreases in recent developments of SRF technologies.  
The same will be equally true in future SRF technologies utilizing alternative materials with smaller $R_{\rm BCS}$~\cite{gurevich1, kubo1, gurevich2, posen_Nb3Sn, kubo2, posen, kubo_SRF2015_multilayer}.

For a long time, shielding cavities from the ambient magnetic field had been the only way to reducing a number of trapped vortices. 
The situation was changed by the recent finding that cool-down conditions affect a number of trapped vortices~\cite{vogt2013, romanenko2014, romanenko2014APL}. 
Romanenko et al. found that cooling down with a larger spatial temperature gradient leads to a better flux expulsion~\cite{romanenko2014APL}. 
They achieved a quality factor $Q_0 \simeq 3\times 10^{11}$ that corresponds to $R_{\rm res} \simeq 1\,{\rm n\Omega}$ under an ambient magnetic field $B_a=1\,\mu{\rm T}$. 
Furthermore, even under a large ambient magnetic field, $B_a=19\,\mu{\rm T}$, they obtained $Q_0 \simeq 6\times 10^{10}$ or $R_{\rm res} \simeq 5\,{\rm n\Omega}$.  
After that, a lot of experimental results that support this finding have been accumulated,  
and a crucial effect of material properties on the flux expulsion has also become recognized: 
a flux expulsion efficiency depends on a cavity material history in addition to the temperature gradient during the cool-down~\cite{posen_SRF2015, posen_TTC2015}.

While substantial developments have been made in experimental studies as above, 
not much theoretical progress has followed on it: 
even the mechanism responsible for a better flux expulsion by the cool-down with a large spatial temperature gradient has not yet been understood. 
In the present paper, 
a model to explain efficient flux expulsions by the cool-down with a large spatial temperature gradient is proposed. 
A number of trapped vortices and resultant $R_{\rm res}$ are naively estimated. 
We find that a number of trapped vortices and $R_{\rm res}$ are proportional to the strength of the ambient magnetic field and the inverse of the temperature gradient 
and that the proportionality constant depends on material properties. 
Then the estimated $R_{\rm res}$ is compared with Romanenko's experiments. 
Effects of material properties on the flux expulsion are also discussed.

%%%%%%%%%%%%%%%%%%%%%%
%%%%%%%%%%%%%%%%%%%%%%
%%%%%%%%%%%%%%%%%%%%%%
%%%%%%%%%%%%%%%%%%%%%%
\section{Model}
%%%%%%%%%%%%%%%%%%%%%%
%%%%%%%%%%%%%%%%%%%%%%
%%%%%%%%%%%%%%%%%%%%%%
%%%%%%%%%%%%%%%%%%%%%%

%
\begin{figure}[tb]
   \begin{center}
   \includegraphics[width=0.6\linewidth]{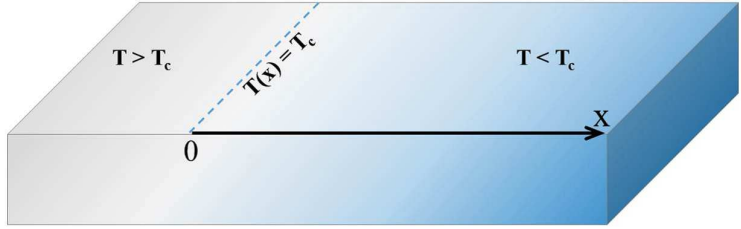}
   \end{center}\vspace{-0.2cm}
   \caption{
A superconducting material cooled down from the right to the left. 
The gray and blue region represent the regions with $T>T_c$ and  $T<T_c$, respectively. 
The origin of the $x$-axis is fixed at the interface of these two regions. 
   }\label{fig1}
\end{figure}

We consider a superconducting material shown in Fig.~\ref{fig1}. 
The gray (blue) region represents a region with $T$ larger (smaller) than the critical temperature $T_c$. 
The origin of the $x$-axis is fixed at the interface of the two regions. 
The material is cooled down from the right to the left: 
$T'(x)\equiv dT/dx<0$ and $dT/dy=dT/dz=0$. 
The material is immersed in an ambient magnetic field, 
where a direction of the ambient field is arbitrary. 
In the following, 
we regard the present system as static and examine the vicinity of $x=0$ or $T= T_c$ in the framework of the Ginzburg$-$Landau (GL) theory~\cite{kubo_SRF2015_flux, kubo_TTC2015}.

%%%%%%%%%%%%%%%%%%%%%%%%%%%%%%%%%%%%%%%%%%%%%%%%%%%%%
\subsection{Phase transition fronts} 
%%%%%%%%%%%%%%%%%%%%%%%%%%%%%%%%%%%%%%%%%%%%%%%%%%%%%

Let us begin with summarizing the temperature dependences of relevant parameters in the GL theory. 
We use the normalized temperature, 
\begin{eqnarray}
\widetilde{T} \equiv \frac{T}{T_c} \,. 
\end{eqnarray}
The coherence length and the penetration depth are given by
$\xi(T) \equiv \xi^* ( 1- \widetilde{T})^{-\frac{1}{2}}$ and $\lambda(T) = \lambda^* ( 1-\widetilde{T} )^{-\frac{1}{2}}$, respectively, 
where $\xi^* \equiv \sqrt{\gamma \hbar^2/|\alpha_0|}$, $\lambda^*=\sqrt{\beta/8\mu_0 e^2 \gamma |\alpha_0|}$, $\hbar$ is the reduced Planck constant, $\mu_0$ is the vacuum permeability, $e$ is the elementary charge, and $\alpha_0$, $\beta$, and $\gamma$ are constants derived from the microscopic theory~\cite{kopnin}.  
A parameter $\kappa \equiv \lambda(T)/ \xi(T) =\lambda^*/ \xi^*$ 
is a temperature independent parameter in the framework of GL theory. 
The upper critical field is given by
$B_{c2}(T) = \phi_0/2\pi \xi(T)^2 = (\phi_0/2\pi \xi^{*2}) ( 1-\widetilde{T})$, 
and the thermodynamic critical field is given by
$B_{c}(T)=B_{c2}(T)/\sqrt{2}\kappa =(\phi_0 /2\sqrt{2}\pi \kappa\xi^{*2}) ( 1-\widetilde{T} )$, 
where $\phi_0=2.07\times 10^{-15}\,{\rm Wb}$ is the flux quantum. 
The lower critical field can also be given by the compact expression, 
$B_{c1}(T) \simeq [ (\ln \kappa + a)/\sqrt{2}\kappa] B_c(T) = [\phi_0 (\ln \kappa + a) /4\pi\kappa^2\xi^{*2}] ( 1-\widetilde{T} )$, 
where $a=0.5+(1+\ln 2)/(2\kappa-\sqrt{2}+2)$ yields good fittings to numerics~\cite{brandt} (e. g. $a=0.5$ for $\kappa \gg 1$ and $a=1.15$ for $\kappa=1$).

\begin{figure}[*t]
   \begin{center}
   \includegraphics[width=1\linewidth]{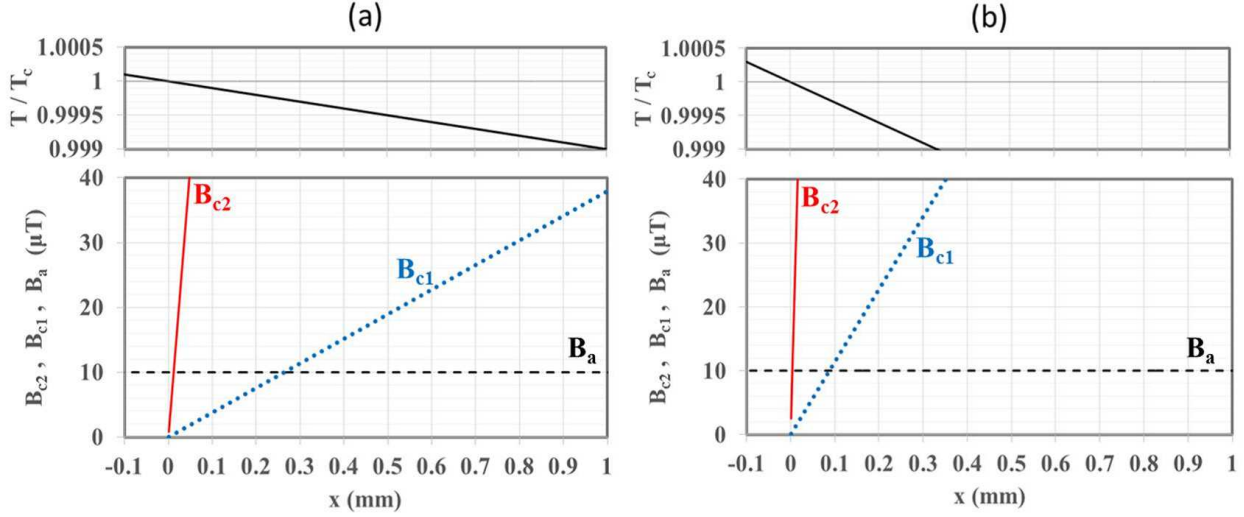}
   \end{center}\vspace{-0.cm}
   \caption{
Examples of temperature and critical field distributions for (a) $|\widetilde{T}'|= |T'|/T_c = 1\,{\rm m^{-1}}$ and (b) $|\widetilde{T}'|= |T'|/T_c = 3\,{\rm m^{-1}}$, 
where $\kappa=5$, $\xi^* = 20\,{\rm nm}$ and an ambient magnetic field $B_a = 10\,\mu{\rm T}$ are assumed as examples. 
   }\label{fig2}
\end{figure}

Since $T$ depends on $x$, 
$T$ dependent parameters also depend on $x$ through a temperature distribution $T(x)$.  
Then the critical fields, $B_{c2}$ and $B_{c1}$, are also given as functions of $x$.  
Since the origin of the $x$-axis is fixed at the position with $T=T_c$, 
the expansion of $T(x)$ in the vicinity of $x=0$ is given by   
\begin{eqnarray}
T(x) =  T_c + T'(0) x 
= T_c \bigl[ 1 - \bigl| \widetilde{T}'(0) \bigr| x \bigr] \,. 
\label{eq:Tdistribution}
\end{eqnarray}
where $\widetilde{T}'(0)$ is defined by $\widetilde{T}'(0)\equiv T'(0)/T_c$, 
and $x\ll |\widetilde{T}'(0)|^{-1}$ is assumed. 
In the following, we abbreviate $T'(0)$ and $\widetilde{T}'(0)$ by $T'$ and $\widetilde{T}'$, respectively. 
By using Eq.~(\ref{eq:Tdistribution}) or $\widetilde{T}(x) = 1-|\widetilde{T}'| x$, 
we find $1-\widetilde{T}(x) = |\widetilde{T}'|x$. 
Then the critical fields are given by
\begin{eqnarray}
&&B_{c2}(T(x)) = \frac{\phi_0}{2\pi \xi^{*2}} \bigl| \widetilde{T}' \bigr| x \,, 
\label{eq:bc2_x} \\
&&B_{c1}(T(x)) = \frac{\phi_0 (\ln \kappa + a) }{4\pi\kappa^2\xi^{*2}} 
\bigl| \widetilde{T}' \bigr| x \,. 
\label{eq:bc1_x}
\end{eqnarray}
Fig.~\ref{fig2}(a) and (b) show examples of $\widetilde{T}(x)$, $B_{c2}(T(x))$ and $B_{c1}(T(x))$ for $|\widetilde{T}'=1\,{\rm m^{-1}}|$ and $3\,{\rm m^{-1}}$, respectively. 
At $x=0$, where $T=T_c$, the critical fields vanish. 
As $x$ increases, $T(x)$ decreases, 
and the critical fields increase with slopes proportional to $|\widetilde{T}'|$.

\begin{figure}[*t]
   \begin{center}
   \includegraphics[width=0.45\linewidth]{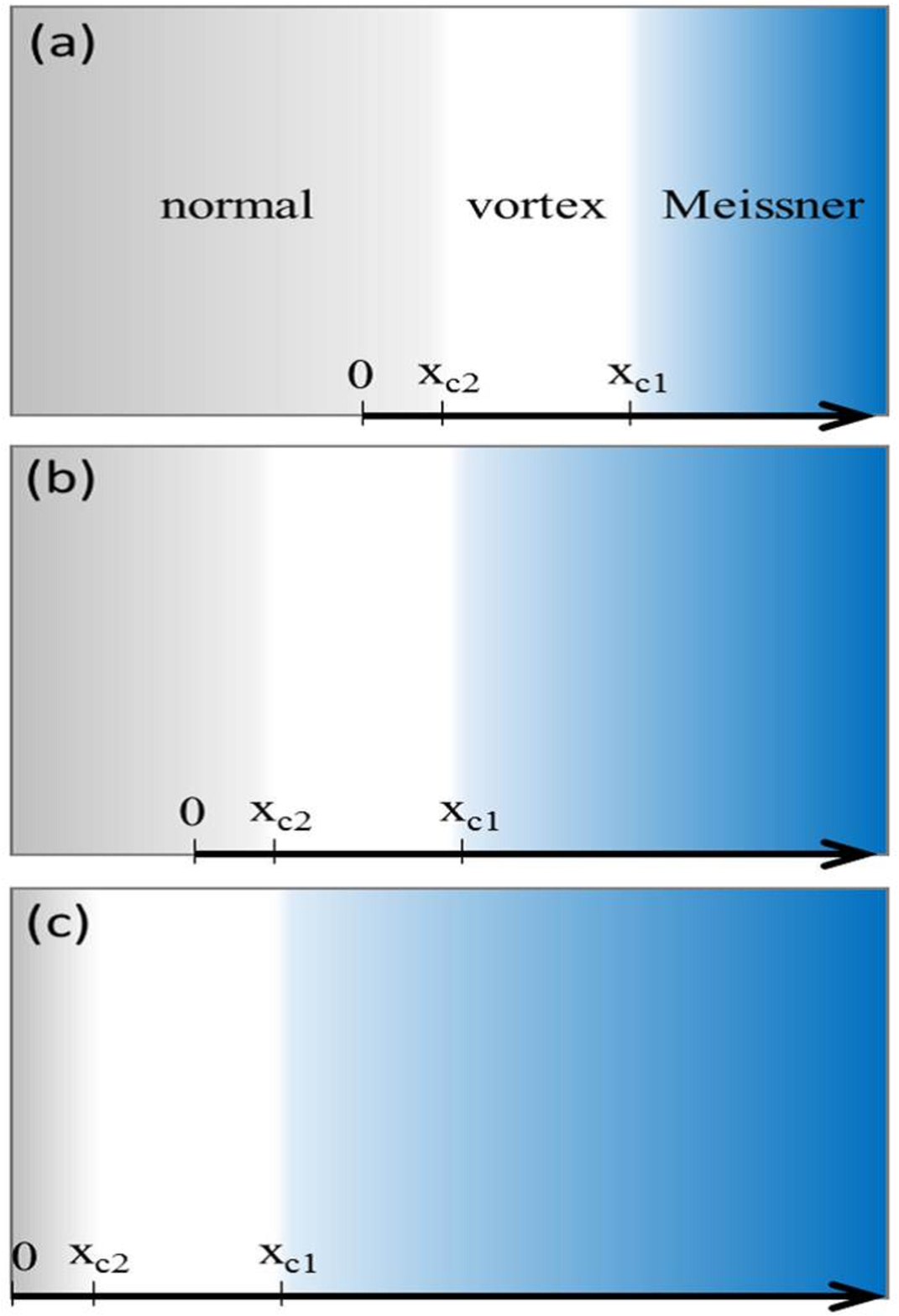}
   \end{center}\vspace{-0.2cm}
   \caption{
Schematic view of the vicinity of the phase transition fronts. 
There exist three domains: 
the normal conducting domain ($x \le x_{c2}$), 
the vortex state domain ($x_{c2} < x \le x_{c1}$), 
and the Meissner state domain ($x > x_{c1}$). 
As the material is cooled down, 
the vortex state domain together with the phase transition fronts sweep the material from the right to the left as shown in (a), (b), and (c). 
   }\label{fig3}
\end{figure}

As shown in Fig.~\ref{fig2}, 
the critical fields are strongly suppressed in the vicinity of $x=0$ or $T=T_c$
and can be smaller than a given ambient magnetic field $B_a$. 
Then there exist two phase transition fronts: 
$x=x_{c2}$ at which $B_{c2}=B_a$ and $x=x_{c1}$ at which $B_{c1}=B_a$. 
The locations of the phase transition fronts are obtained by substituting $x=x_{c2}$ ($x=x_{c1}$) and $B_{c2}=B_a$ ($B_{c1}=B_a$) into Eq.~(\ref{eq:bc2_x}) (Eq.~(\ref{eq:bc1_x})). 
Assuming that $B_{a}$ is spatially uniform, we find 
\begin{eqnarray}
&&x_{c2} = 2\pi \xi^{*2} \frac{B_a}{\phi_0} \bigl| \widetilde{T}'\bigr|^{-1} \,, 
\label{eq:xc2} \\
&&x_{c1} = \frac{4\pi \kappa^2 \xi^{*2}}{\ln \kappa + a} \frac{B_a}{\phi_0} 
\bigl| \widetilde{T}' \bigr|^{-1} \,. 
\label{eq:xc1} 
\end{eqnarray}
As shown in Fig.~\ref{fig3}(a), 
the material is in the normal conducting state and the Meissner state at $x \le x_{c2}$ and $x > x_{c1}$, respectively, 
and the region between $x_{c2}$ and $x_{c1}$ is in the vortex state, 
whose thickness is given by 
\begin{eqnarray}
\delta x \equiv x_{c1}-x_{c2} 
= 4\pi \xi^{*2} f_{-}(\kappa)
\frac{B_a}{\phi_0} \bigl| \widetilde{T}' \bigr|^{-1} 
\,,  \label{eq:deltax} 
\end{eqnarray}
where $f_{-} (\kappa) \equiv \kappa^2/(\ln\kappa +a) - 1/2$. 
Let us check concrete values of $x_{c2}$, $x_{c1}$, and $\delta x$. 
Taking the parameters of Fig.~\ref{fig2} for example, 
we obtain (a) $x_{c2}=12\,{\rm \mu m}$, $x_{c1}=2.7\times 10^2\,{\rm \mu m}$, and $\delta x=2.6\times 10^2\,{\rm \mu m}$; 
(b) $x_{c2}=4\,{\rm \mu m}$, $x_{c1}=89\,{\rm \mu m}$, and $\delta x=85\,{\rm \mu m}$. 
The behaviors of Eqs.~(\ref{eq:xc2})-(\ref{eq:deltax}) as functions of $B_a$ and $|\widetilde{T}'|$ can be understood as follows.  
An increase of $B_a$, which corresponds to pushing up the horizontal dashed line, 
increases $x_{c2}$ and $x_{c1}$ and expands $\delta x$.  
An increase of $| \widetilde{T}' |$ makes the slopes of the red solid line and the blue dotted line steeper and allows $B_{c2}$ and $B_{c1}$ make recoveries in shorter lengths. 
Then $x_{c2}$ and $x_{c1}$ decrease, and $\delta x$ is compressed.

All the above calculations are based on the GL theory and are valid near $T\simeq T_c$ or $x \ll |\widetilde{T}'|^{-1}$. 
Thus Eqs.~(\ref{eq:xc2}) and (\ref{eq:xc1}) are valid only when $B_a/\phi_0 \ll (1/2\pi\xi^{*2})$ and $B_a/\phi_0 \ll \kappa^{-2}(\ln\kappa +a)/ 4\pi \xi^{*2}$, respectively. 
These conditions are usually satisfied as long as a typical ambient magnetic field $\lesssim \mathcal{O}(10)\,\mu{\rm T}$ is assumed. 
If this is not the case, the phase transition fronts shift to a low temperature region due to a large $B_a$, 
the assumption $T\simeq T_c$ ceases to be valid, 
and the GL framework becomes no longer applicable.

In the vortex state domain, vortices penetrate the material, 
whose core sizes roughly correspond to the coherence length $\xi(T)$ and thus depend on $x$ through the temperature distribution $T(x)$. 
Substituting Eq.~(\ref{eq:Tdistribution}) or $(1-\widetilde{T}(x))^{-\frac{1}{2}}=x^{-\frac{1}{2}} |\widetilde{T}'|^{-\frac{1}{2}}$ into $\xi = \xi^* (1-\widetilde{T})^{-\frac{1}{2}}$, 
we obtain
\begin{eqnarray}
\xi(T(x))=\frac{\xi^*}{\sqrt{x}} \bigl| \widetilde{T}' \bigr|^{-\frac{1}{2}} \,,  
\label{eq:xi_x}  
\end{eqnarray}
which varies from a maximum at the hottest region $x=x_{c2}$ to a minimum at the coldest region $x=x_{c1}$ as $x$ increases. 
The maximum and the minimum are given by $\xi|_{x_{c2}}=(1/\sqrt{2\pi}) \sqrt{\phi_0/B_a}$ and $\xi|_{x_{c1}}=\sqrt{(\ln \kappa + a)/4\pi\kappa^2} \sqrt{\phi_0/B_a}$, respectively. 
Taking the parameter set of Fig.~\ref{fig2} for example, 
we obtain $\xi|_{x_{c2}}=6\,{\rm \mu m}$ and $\xi|_{x_{c1}}=1\,{\rm \mu m}$. 
In general, the ratio of the maximum to the minimum, 
$\xi|_{x_{c2}}/\xi|_{x_{c1}} = \sqrt{x_{c1}/x_{c2}}=\kappa\sqrt{2/(\ln\kappa + a)}$, 
is not so large as long as a material with $\kappa \lesssim \mathcal{O}(10^2)$ is assumed.  
We can define a representative value of $\xi$ in the vortex state domain as
\begin{eqnarray}
\bar{\xi} \equiv \xi(T(\bar{x})) = \frac{1}{\sqrt{2\pi f_+(\kappa)}} \sqrt{\frac{\phi_0}{B_a}}
\,, \label{eq:xi_xbar}  
\end{eqnarray}
where $\bar{x} \equiv (x_{c1}+x_{c2})/2 = 2\pi\xi^{*2} f_{+}(\kappa) (B_a/\phi_0) |\widetilde{T}'|^{-1}$ 
and $f_{+} (\kappa) \equiv \kappa^2/(\ln\kappa +a) + 1/2$. 
Note that, as $B_a$ increases, the vortex state domain shifts to a colder region, 
and $\bar{\xi}$ decreases.

%%%%%%%%%%%%%%%%%%%%%%%%%%%%%%%%%%%%%%%%%%%%%%%
\subsection{Flux trapping and residual resistance}
%%%%%%%%%%%%%%%%%%%%%%%%%%%%%%%%%%%%%%%%%%%%%%%

As the material is cooled down [see Fig.~\ref{fig3}(a), (b), and (c)], 
vortices contained in the vortex state domain are transported together with the propagations of the phase transition fronts and trapped by pinning centers such as normal conducting precipitates, grain boundaries, etc. 
For the case of Nb cavity, 
dominant contributions to a number of trapped vortices $N_{\rm trap}$ are thought to come from those trapped by non-superconducting precipitates. 
For a calculation of a pinning force due to a non-superconducting nano precipitate,  
the GL framework that correctly include an impurity potential is available~\cite{thuneberg, friesen}.  
Then $N_{\rm trap}$ may be evaluated by three dimensional simulations of a many body system that consists of a number of pinning centers and moving vortices in the GL theory. 
Instead of such tough simulations, here we roughly estimate $N_{\rm trap}$ in the following.

\begin{figure}[*t]
   \begin{center}
   \includegraphics[width=0.8\linewidth]{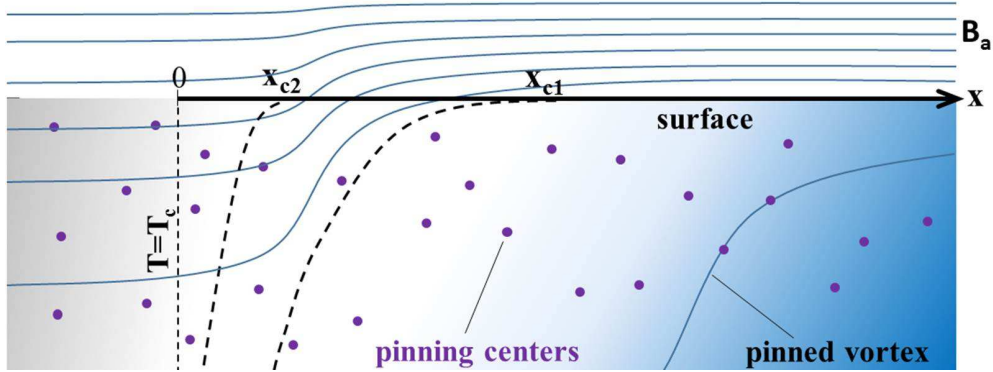}
   \end{center}\vspace{-0.1cm}
   \caption{
A schematic view of the vicinity of the phase transition fronts with an ambient magnetic field parallel to the surface. 
The region between the two dashed curves labeled by $x_{c2}$ and $x_{c1}$ corresponds to the vortex state domain. 
The dots represent pinning centers. 
   }\label{fig4}
\end{figure}

For a while, we assume a direction of an ambient magnetic field is parallel to the $x$-axis for simplicity (see Fig.~\ref{fig4}). 
This assumption corresponds to the field configuration at the equator of cavity in Ref.~\cite{romanenko2014APL}. 
In order to estimate $N_{\rm trap}$, we use an analogy with a beam-target collision event, 
where pinning centers and the vortex state domain correspond to beams and a target, respectively. 
Since $N_{\rm trap}$ is expected to be proportional to a number of collision events. 
we obtain
\begin{eqnarray}
N_{\rm trap} \propto N_{\rm pin} N_{\phi} \sigma \,, 
\label{eq:Ntrap1}
\end{eqnarray}
where $N_{\rm pin}$ is a number of relevant pinning centers, 
$N_{\phi}$ is a number of vortices contained in the vortex state domain, 
and $\sigma$ is a  reaction cross-section. 
First, $N_{\rm pin}$ is not a total number of pinning centers but a number of relevant pinning centers that can pin vortices against unpinning forces. 
Thus, for example, $N_{\rm pin}$ does not include a number of pinning centers weaker than the thermal force.~\footnote{ 
A pinning force $f_{\rm pin}(T)$ approaches zero as $T\to T_c$. 
The thermal force~\cite{huebener} given by $f_T(T) = -s^* \nabla T$ with $s^* \simeq - (\phi_0/\mu_0) \partial B_{c1}(T)/\partial T$ also approaches zero as $T\to T_c$. 
Then the ratio $|f_{\rm pin} / f_T|$ diverges (vanishes) when $|f_{\rm pin}|$ approaches zero slower (faster) than $|f_T|$ as $T\to T_c$. 
In the vicinity of the phase transition fronts, where $T\simeq T_c$, 
the ratio $|f_{\rm pin} / f_T|$ is strongly enhanced or suppressed; 
pinned vortices remain trapped for the former case and are flushed out for the latter case~\cite{gurevich_ciovati, gurevich_TTC2015}. 
$N_{\rm pin}$ includes a number of pinning center that correspond to the former case, but does not include those correspond to the latter case.} 
Next, $N_{\phi}$ is proportional to a product of a vortex density $\sim B_a/\phi_0$ and a thickness $\delta x$: 
\begin{eqnarray}
N_{\phi} \propto B_a \delta x \propto B_a^2 \bigl| \widetilde{T}' \bigr|^{-1} 
\,. 
\label{eq:Nphi}
\end{eqnarray}
It should be noted that, 
even if some vortices are trapped while the vortex state domain sweeps the material, 
$N_{\phi}$ remains almost constant, 
because vortices are supplied from the ambient field. 
Finally, we roughly evaluate $\sigma$. 
Since $f_{\rm pin}$ can reach a distance of the order of $\xi$~\cite{matsushita}, 
$\sigma$ can be naively estimated as
\begin{eqnarray}
\sigma \propto  \bar{\xi}^2 \propto B_a^{-1} 
\,, \label{eq:sigma}
\end{eqnarray}
where $\bar{\xi}\equiv \xi(\bar{x})$ is used as a representative value of $\xi$ in the vortex state domain.  
Note that, independently of a choice of $x$ to evaluate $\xi$, we obtain the same proportionality relation as Eq.~(\ref{eq:sigma}) (e. g. $\xi|_{x_{c2}}$ and $\xi|_{x_{c1}}$ are also proportional to $1/\sqrt{B_a}$ and lead to the same result). 
Substituting Eqs.~(\ref{eq:Nphi}) and (\ref{eq:sigma}) into Eq.~(\ref{eq:Ntrap1}), 
we find  
\begin{eqnarray}
N_{\rm trap} = \mathcal{A_M} B_a \bigl| \widetilde{T}' \bigr|^{-1} \,,
\label{eq:Ntrap0}
\end{eqnarray}
where $\mathcal{A_M}$ is a material dependent parameter proportional to $N_{\rm pin}$ and a function of $\kappa$ and $\xi^*$. 
The existence of the factor $|\widetilde{T}'|^{-1}$ can be understood as follows.  
As a temperature gradient increases, a thickness of the vortex state domain decreases [see Eq.~(\ref{eq:deltax})], 
and a number of vortices contained in the vortex state domain decreases. 
Then a reaction probability decreases, 
and a number of trapped vortices, $N_{\rm trap}$, decreases.  
Note that the present model ceases to be valid when $|\widetilde{T}'|$ is so large that the thickness of the vortex state domain $\delta x$ is smaller than the order of $\xi$: 
there exists a cutoff at a large $|\widetilde{T}'|$. 
There is no reason that we regard a contribution above the cutoff vanishes. 
Thus, in general, we have
\begin{eqnarray}
N_{\rm trap} = \mathcal{A_M} B_a \bigl( \bigl| \widetilde{T}' \bigr|^{-1} + \mathcal{D_M} \bigr) \,,
\label{eq:Ntrap}
\end{eqnarray}
where $\mathcal{D_M}$ is a contribution that remains above the cutoff of the present model and generally depends on a material property. 
The concrete form of $\mathcal{D_M}$ is expected to be revealed by rigorous analyses using the time dependent GL equation. 
When $|\widetilde{T}'|$ is too small, the unitarity is broken: a total fluxes trapped by the material exceeds the original total ambient fluxes around the material. 
Thus a cutoff also exists at a small $|\widetilde{T}'|$.

Oscillating trapped vortices under the RF field contribute to the RF dissipation $P_{v}$ and $R_{\rm res}$, 
where $R_{\rm res}\equiv  2P_{v}/H_{\rm RF}^2$ and $H_{\rm RF}$ is the RF magnetic field. 
Each trapped vortex individually contributes to $P_{v}$, 
and $P_{v}\propto N_{\rm trap}$. 
Then we obtain $R_{\rm res} = 2P_{v}/H_{\rm RF}^2 \propto N_{\rm trap}$ or 
\begin{eqnarray}
R_{\rm res} = \mathcal{C_M} B_a \bigl( \bigl| \widetilde{T}' \bigr|^{-1} + \mathcal{D_M} \bigr) \,,
\label{eq:Rres}
\end{eqnarray}
where $\mathcal{C_M}$ is a free parameter. 
In order to get a more concrete form of $\mathcal{C_M}$, 
a model that relates a vortex dynamics with $R_{\rm res}$ is necessary. 
We use the result of Ref.~\cite{gurevich_ciovati}.   
By using our notation, their result can be written as $R_{\rm res} \propto  r N_{\rm trap}$, 
where $r$ is called the sensitivity and corresponds to $R_i$ given by Eq.~(25) of Ref.~\cite{gurevich_ciovati} normalized by a trapped flux density $B_0$. 
Then we find $\mathcal{C_M}\propto \mathcal{A_M} r$.

An ambient magnetic field with an arbitrary direction also leads to the same field configuration as Fig.~\ref{fig4},  
because the ambient magnetic field is bent by the diamagnetic property of the Meissner state domain and becomes parallel to the $x$-axis in the vicinity of the vortex state domain. 
Then $N_{\rm trap}$ and $R_{\rm res}$ can be evaluated in much the same way as the above and are given by Eq.~(\ref{eq:Ntrap}) and Eq.~(\ref{eq:Rres}), respectively.

%%%%%%%%%%%%%%%%%%%%%%%%%%%%%%%%%
%%%%%%%%%%%%%%%%%%%%%%%%%%%%%%%%%
%%%%%%%%%%%%%%%%%%%%%%%%%%%%%%%%%
\section{Discussion}
%%%%%%%%%%%%%%%%%%%%%%%%%%%%%%%%%
%%%%%%%%%%%%%%%%%%%%%%%%%%%%%%%%%
%%%%%%%%%%%%%%%%%%%%%%%%%%%%%%%%%

%
\begin{figure}[t]
   \begin{center}
   \includegraphics[width=0.6\linewidth]{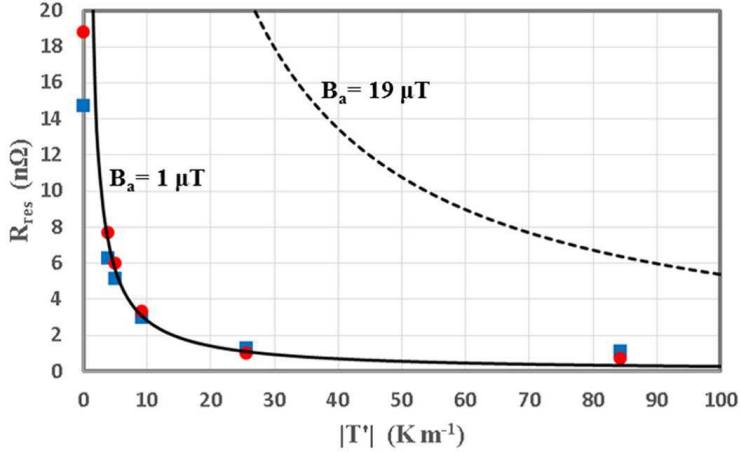}
   \end{center}\vspace{-0.2cm}
   \caption{
$R_{\rm res}$ as functions of $|T'|$. 
The blue squares and the red filled circles represent experimental data read from Ref.~\cite{romanenko2014APL}, 
where a distance between the two sensors at the equator and the iris is assumed to be $0.1\,{\rm m}$. 
The curves represent Eq.~(\ref{eq:Rres}) with $\mathcal{C_M}=3.08\,{\rm n\Omega \,m^{-1} \,\mu T^{-1}}$. 
The solid and dashed curves correspond to $B_a =1\,\mu{\rm T}$ and $19\,\mu{\rm T}$, respectively. 
   }\label{fig5}
\end{figure}

Let us compare our result with the experimental result obtained by Romanenko~\cite{romanenko2014APL}. 
Here we assume $\mathcal{D_M}$ is so small and neglect its contribution for simplicity.  
Fig.~\ref{fig5} shows $R_{\rm res}$ as a function of $|T'|\,(=T_c |\widetilde{T}'|)$. 
The blue squares and red filled circles represent experimental data for $B_a=1\,{\rm \mu T}$, 
where the temperature differences in Ref.~\cite{romanenko2014APL} are translated into the temperature gradients by assuming the distance between the two temperature sensors at the equator and the iris $\simeq 0.1\,{\rm m}$. 
As $|T'|$ increases, $R_{\rm res}$ decreases.  
The solid curve represents Eq.~(\ref{eq:Rres}) with $\mathcal{C_M}=3.08\,{\rm n\Omega \,m^{-1} \,\mu T^{-1}}$ and $B_a=1\,{\rm \mu T}$, 
which agrees well with the experimental data. 
By using Eq.~(\ref{eq:Rres}) with the same $\mathcal{C_M}$ as the above, 
we can predict $R_{\rm res}$ of the same cavity under a different $B_a$. 
The dashed curve represents $R_{\rm res}$ for $B_a=19\,\mu {\rm T}$ calculated by using Eq.~(\ref{eq:Rres}),  
which shows $R_{\rm res}\simeq 5\,{\rm n\Omega}$ can be achieved by a large $|T'|\simeq 1\times 10^2\,{\rm K/m}$ even under such a strong ambient magnetic field. 
This prediction is consistent with Ref.~\cite{romanenko2014APL}: 
they achieved $Q_0 \simeq 6 \times 10^{10}$ or $R_{\rm res} \simeq 5\,{\rm n\Omega}$ under $B_a=19\,\mu {\rm T}$.

Material properties affect $N_{\rm trap}$ through the material dependent parameters  $\mathcal{A_M}$ and $\mathcal{D_M}$. 
Not only cool-down conditions but also parameters inherent to a material have crucial effects on a number of trapped vortices or the flux expulsion efficiency (e. g. since $\mathcal{A_M}$ is proportional to $N_{\rm pin}$, $N_{\rm trap}$ is also proportional to $N_{\rm pin}$).  
Then, even if we successfully generate a large $|\widetilde{T}'|$ and achieve an almost complete flux expulsion for a given cavity, 
the same $|\widetilde{T}'|$ does not necessarily lead to a satisfying flux expulsion for other cavities.  
$\mathcal{A_M}$ and $\mathcal{D_M}$ must be reduced to improve a flux expulsion efficiency. 
Since the form of $\mathcal{D_M}$ is completely unknown in the present model, 
we can say nothing about a way to reducing $\mathcal{D_M}$.  
However, a reduction of $\mathcal{A_M}$ can be achieved by decreasing $N_{\rm pin}$. 
A high temperature treatment to disperse impurity precipitates is expected to be effective in a reduction of $N_{\rm pin}$. 
A pure material with a large residual resistance ratio (RRR) is also expected to be effective. 
Use of ingot Nb materials may also be effective, 
because total amount of non-superconducting impurities in grain boundaries is expected to be smaller than polycrystalline Nb sheets, and the production processes of Nb discs from ingots include smaller possibilities of involving non-superconducting impurities than those of polycrystalline Nb sheets: 
forging, rolling, and many other steps are no longer necessary for use of ingot Nb materials~\cite{saito_SRF2009, umezawa_TTC2015}. 
Recent experimental results are consistent with these statements~\cite{posen_SRF2015, posen_TTC2015}.

The sensitivity of $R_{\rm res}$ to trapped flux density should also be touched on here. 
It is well known that the sensitivity is affected by material properties~\cite{saito}.    
Recently, the mean free path dependence of the sensitivity is studied strenuously~\cite{martinello_SRF2015, martinello_TTC2015, gonnella_SRF2015, gonnella_TTC2015}. 
In the present model, the sensitivity is given by $r$ identical to that of Ref.~\cite{gurevich_ciovati} and depends not only on the mean free path but also on the mean spacing between pinning centers that is generally independent of the mean free path~\cite{gurevich_ciovati} (see also Ref.~\cite{gurevich_TTC2015}).  
Then, in order for looking at the mean free path dependence of the sensitivity in experiments, 
the mean spacing between pinning centers must be common among materials.

A direction of the ambient field and a global cool-down direction relative to a cavity direction can also affect a number of trapped vortices as pointed in Ref.~\cite{martinello}. 
Taking the horizontal cavity cooled down from the bottom equator in a vertical ambient field for example, 
some field lines are finally encircled by superconducting regions and highly concentrated at the top equator, which are trapped as bundle of vortices and contribute to $R_{\rm res}$ (see also Ref.~\cite{kubo_TTC2015}). 
The present model, which focuses only local dynamics in the vicinity of the phase transition fronts, 
does not take into account this tricky phenomenon caused by the global configuration. 
The present model is applicable only when this tricky phenomenon is absent. 
This problem is expected to be avoided by changing a global cool-down direction relative to a cavity direction and shifting the concentration region of expelled field lines from the equator to other regions (e. g. installing heaters at the iris of each cell for the case of the horizontal cavity cooled down from the bottom).

Alternative materials~\cite{gurevich1, kubo1, gurevich2, posen_Nb3Sn, kubo2, posen, kubo_SRF2015_multilayer} are often used in the form of a film similar to or thinner than a typical size of vortices in the vortex state domain $\sim \mathcal{O}(1)\,{\rm \mu m}$, 
where the present model is no longer valid. 
The same goes for the traditional Nb film technology~\cite{benvenuti}. 
Furthermore, in the derivation of $N_{\rm trap}$, pinning centers are assumed to be zero-dimensional defects. 
However, pinning centers can not necessarily be regarded as zero-dimensional. 
Taking ${\rm Nb_3 Sn}$ for example, the major pinning centers are grain boundaries~\cite{matsushita}, 
which are two-dimensional objects, and can not be treated in the present model. 
Evaluations of $N_{\rm trap}$ and $R_{\rm res}$ for materials other than bulk Nb are the future challenges.

%%%%%%%%%%%%%%%%%%%%%%%%%%%%%%%%%
%%%%%%%%%%%%%%%%%%%%%%%%%%%%%%%%%
%%%%%%%%%%%%%%%%%%%%%%%%%%%%%%%%%
\section{Summary}
%%%%%%%%%%%%%%%%%%%%%%%%%%%%%%%%%
%%%%%%%%%%%%%%%%%%%%%%%%%%%%%%%%%
%%%%%%%%%%%%%%%%%%%%%%%%%%%%%%%%%

A magnetic flux is trapped as quantized vortices during the cool-down of a superconductor through its critical temperature $T_c$, 
which yields additional dissipation and contributes to the residual resistance $R_{\rm res}$. 
Recently, cooling down with a large spatial temperature gradient attracts much attention for successful reductions of trapped vortices. 
Its mechanism, however, has not been well understood.

In the present paper, 
a model to explain the efficient flux expulsions by the cool-down with a large spatial temperature gradient was proposed. 
We considered the simple model shown in Fig.~\ref{fig1}. 
In the vicinity of the region with $T \simeq T_c$, 
$B_{c1}$ and $B_{c2}$ are strongly suppressed and can be smaller than the ambient magnetic field, $B_a$, as shown in Fig.~\ref{fig2}. 
There exist two phase transition fronts: 
$x=x_{c2}$ at which $B_{c2}=B_a$ and $x=x_{c1}$ at which $B_{c1}=B_a$. 
The region $x_{c2} < x \le x_{c1}$ is the vortex state domain. 
As the material is cooled down, 
the phase transition fronts together with the vortex state domain sweep the material as shown in Fig.~\ref{fig3}. 
In this process, vortices contained in the vortex state domain are trapped by pinning centers that randomly distribute in the material. 
We naively estimated a number of trapped vortices $N_{\rm trap}$ by using the analogy with the beam-target collision event instead of tackling three dimensional simulations of the GL theory. 
$N_{\rm trap}$ is given by Eq.~(\ref{eq:Ntrap}).  
By using this together with the previous result in Ref.~\cite{gurevich_ciovati}, 
we evaluated $R_{\rm res}$ and obtained Eq.~(\ref{eq:Rres}).

The obtained residual resistance $R_{\rm res}$ was compared with experiments;
it agreed well with the experimental data as shown in Fig.~\ref{fig5}. 
The material dependence of a number of trapped vortices $N_{\rm trap}$ were also discussed qualitatively; not only the cool-down conditions but also material properties, such as a number of relevant pinning centers, crucially affects $N_{\rm trap}$.

\end{document}